# Prospects in MPGDs development for neutron detection

Bruno Guerard (ILL), Richard Hall-Wilton (ESS), Fabrizio Murtas (INFN & CERN)

## Introduction

The market of neutron detectors has increased significantly during the last decade in two domains: instrumentation for Neutron Scattering Science, and prevention against nuclear terrorism. Before the emerging of the so-called "$^3$He shortage crisis", detection systems used in portal monitors to detect fissile elements were based mainly on $^3$He proportional counters, whereas linear $^3$He PSDs (Position Sensitive Detectors), $^3$He MWPCs, and $^6$Li scintillators were the most current techniques for scientific applications. Two large-scale neutron facilities, SNS in the US and J-PARC in Japan, have recently started their operation, and the future ESS (European Spallation Source) will produce first neutrons in 2019-2020. Detectors with better performance are urgently needed to take full benefit of the high intensity neutron beams produced by these sources. An additional constraint comes from the fact that the volume of $^3$He available is by far insufficient to cope with the demand for large area detectors, and the cost of this gas has increased considerably.

Compared to Multi Wires Proportional Chambers (MWPC), Micro-Pattern Gas Detectors (MPGD) used in HEP to detect MIPs offer better spatial resolution, counting rate capability, and radiation hardness; their fabrication is also more reproducible. Provided similar advantages are applicable to detect neutrons, MPGDs might contribute significantly to the development of neutron scientific instrumentation. In order to evaluate the prospects of neutron MPGDs, it is worth knowing the applications which would benefit from a gain in performance, and if they offer a competitive alternative to conventional $^3$He detectors. These questions have been at the focus of the workshop "Neutron Detection with Micro-Pattern Gaseous Detectors" organized by RD51 in collaboration with HEPTech, which took place at CERN on October 14-15, 2013. The goal of this workshop was to help disseminating MPGD technologies beyond High Energy Physics, and to give the possibility to academic institutions, potential users and industry to meet together. 26 speakers gave presentations on the following topics:

- MPGD state of the art in HEP
- Non-HEP applications
- MPGDs fabrication process
- Neutron detectors applications
- Design of neutron gas detectors

Half of the presentations were focused on neutron scattering science applications, providing a complete overview on the demand for high performance neutron detectors and for $^3$He alternatives; other applications presented are in the fields of nuclear physics, reactor control, well logging, and homeland security. The workshop ended with a round table, animated by a panel of 4 detector physicists representing CERN, ESS, INFN, ILL, and the Centronic Company. This round table was fruitful to answer questions about the specificities of neutron detectors, and to discuss new ideas based on MPGDs. The mutual interest between the 2 communities, HEP and neutron scattering science, to work together on the development of new neutron detectors based on MPGDs, was tangible.

This article, written by the participants of the round table, starts with a short summary of the state of the art of MPGD techniques for HEP; then specificities in the design of neutron detector are described; the consequence of the $^3$He shortage are briefly discussed; requirements for neutron scattering science at current facilities and at the future ESS are described; finally, we give some recommendations about possible directions where we believe the development of neutron MPGDs is of particular interest for possible use on neutron scattering instruments. We put some references essentially related to the presentations of the workshop that hopefully will help the readers to obtain more detailed information regarding the MPGD technology and results obtained so far on neutron detections. We apologize for the long delay between the workshop and the delivery of this summary, and for our oriented angle of view which is strongly influenced by our background in gas detectors and in thermal neutron detectors. Despite their high representation in neutron detectors techniques, scintillators and fast neutron detectors were only briefly presented during the workshop and are not discussed in detail here.

## MPGDs state of the art

The rate capability of a MWPC is limited to a value of $10^4$ Hz/mm$^2$; this is not acceptable for modern physics experiments, where the interaction rate can be even two orders of magnitude ($10^6$ Hz/mm2) higher. The development of MPGDs took off in the 1990s mainly as a way to achieve a higher rate capability than the one of MWPCs. The first micropattern gaseous detector was the Micro Strip Gas Chamber (MSGC) introduced in 1988 by Anton Oed for neutron scattering science [1]. The current trend on MPGDs in HEP focuses mainly on GEMs (Gas Electron Multiplier), Micromegas and their derivatives. GEM and Micromegas are currently used for HEP experiment (e.g COMPASS and LHCb at CERN) and new detectors upgrades with these technologies are foreseen for ATLAS and CMS experiments.

Invented by Fabio Sauli in 1997, the Gas Electron Multiplier [2] consists in a thin (50µm) insulating foil copper-clad on both sides. This foil is perforated and a high density, regular matrix of holes (around 100 per square millimeter) is created. Typically, the distance between holes (pitch) is 140 µm and their diameter is about 70 µm. The mesh is realized by exploiting conventional photolithographic methods as used for the fabrication of multi-layer boards. Nowadays the most exploited technique is the Triple GEM structure where three GEM foils are sandwiched between a cathode and an anode. Due to the separation from the readout structure, possible discharges do not directly impact the front-end electronics, making the detector more discharge tolerant. Multiple GEM foils can be cascaded in order to achieve higher gain while keeping a lower voltage on each GEM and decreasing the discharge probability. The triple GEM has now become a standard in HEP, used in many high rate applications. In addition, GEMs have a natural property of suppressing ion feedback, making them very interesting for Time Projection Chambers (TPCs) end-caps.

Another detector structure developed in the same period by I. Giomataris is the micromesh gaseous detector, or Micromegas [3]. A Micromegas detector is a miniaturized version of a very asymmetric two-stage parallel plate detector. A micromesh separates the conversion space, whose dimension typically ranges from 2 mm up to 10 mm, from a small amplification gap that can be as small as 50 µm. This configuration allows, by applying reasonable potentials to the electrodes, to obtain a very high electric field in the amplification region and a quite low electric field in the drift region ($E_d$). Primary electrons are produced in the conversion region and are transported into the amplification region where the multiplication avalanche is created. The ratio between the electric field in the amplification gap and the one in the conversion gap can be tuned to large values, as it is required for an optimal functioning of the device. Besides, such a high ratio is also required in order to catch the ions in the small amplification gap: under the action of the high electric field, the ion cloud is quickly collected on the micromesh and only a small part of it, inversely proportional to the $E_a/E_d$ ratio, escapes towards the conversion region.

GEM and Micromegas have also been developed with B and $B_4C$ convertor films for neutron physics, for example they have being used in the n-TOF project to measure neutron interaction cross sections and the beam size, and borated multiple-GEM detectors are being employed in Dubna, in the so-called CASCADE configurations, on the EPSILON-MDS diffractometer and at FRM II on two instruments, RESEDA as well as MIRA in an resonance spin-echo configuration. GEM-based fast and thermal neutron beam monitors have also been used to characterize the ISIS neutron beams.

These detectors offer intrinsic advantages with respect to MWPC for neutron instrumentation and some talks during the workshop have shown promising results on detection efficiency, rate capability, gamma discrimination and spatial uniformity.

Both GEMs (in the triple GEM configuration) and Micromegas can reach gains of $10^5$ for Minimum Ionizing Particles (MIP) but, if GEM have a discharge probability of $2 \times 10^{-3}$ at gain $10^5$[4], Micromegas show a discharge rate of 80 Hz at a gain of $10^4$ [5]. In order to solve this discharge problem, a big R&D effort was recently performed introducing resistive layers in the Micromegas read-out [6]. It is important to note that the low amplification gain required to detect neutrons strongly reduces this problem. The maximum sensitive area without dead space will also be an important criteria, as well as high stability and reliability in time.

## Specificities of neutron gas detectors

There are basically two types of interaction by which neutrons can be detected: (n,p) elastic scattering on Hydrogen, and nuclear capture. In the first interaction, the kinetic energy of neutrons is transferred to protons in a Hydrogen-rich materials, and the proton recoil is measured. This reaction is exploited to detect fast neutrons for different applications like hadron therapy, diagnostics in controlled thermonuclear fusion, or homeland security to localize sources of neutrons. Neutron detectors based on the proton recoil measurement can provide information on the energy, and on the trajectory of impinging neutrons; another advantage is that the prompt interaction can be exploited for high count rate detectors. Triple GEM are intensively studied for this type of applications at INFN, Frascati and Milano Bicocca. A disadvantage of the proton recoil technique, which precludes its use in neutron scattering science, is the very low detection efficiency (of the order of $10^{-4} - 10^{-5}$) it provides, but good enough for neutron beam monitors[7-8-9]. The second technique to detect fast neutrons is to first reduce their energy by multiple elastic interactions in a H-rich medium, and then to detect the thermalized neutrons by nuclear capture interaction in a thermal neutron detector. It is possible to determine the energy spectrum of a neutron beam with a system of Bonner spheres with different diameters. Since the thermalized neutron is generally detected at a long distance from the first (n,n') interaction, it is not possible to localize the neutrons with good precision. Another limitation comes from the dead time induced by the thermalization process. Proportional counter tubes are well suited for this application.

In order to match the energy range required to probe materials, neutron produced in large scale facilities are first thermalized before being delivered to the instruments. Thus, only thermal neutron detectors are needed. BF3 proportional counter tubes used originally have been replaced by $^3$He position sensitive detectors in most of the neutron instruments. A general, and very simplified description of a neutron PSDs principle helps to illustrate some of their specificities and to define their requirements in the next section.

A thermal neutron gas detector is composed of:
1) a gas vessel with an entrance window transparent to neutrons
2) internal electrodes polarized to appropriate voltage to collect and amplify the charge signal produced by the interactions of neutrons,
3) processing electronics to record the position and the detection time for each neutron

The material of the entrance window is generally made of a few mm of Aluminium, or a few tenths of mm of stainless steel. Organic materials, rich in H, are forbidden due to the scattering of neutrons.

Three elements are needed to convert thermal neutrons into ionizing particles (convertor), control the range of the ion track (stopping gas), and enable gas amplification process (quenching). One element might fulfill one, two, or even three of these functions. This is the case for example for BF3 detectors, where the Boron is isotopically enriched in $^{10}$B to maximize the detection efficiency. The $BF_3$ molecule is a very good quencher with good stopping power, but this gas is electronegative and toxic and has a 30% lower cross section compared to $^3$He; due to its electro-negativity and high quenching factor, the maximum pressure acceptable for a $BF_3$ proportional tube is typically 2 bar for 2.5 cm diameter tubes. It follows that three $BF_3$ tubes are needed to replace one $^3$He tube to get the same detection efficiency. Despite the increase of complexity, $BF_3$ detectors is a cost effective solution, immediately transferable to the industry for large scale production, but the risk due to its toxicity is not acceptable.

More often, the conversion function is separated from the two other functions, as it is the case for $^3$He based detectors, and for gas detectors based on thin convertor films. For $^3$He detectors, the gas vessel must be perfectly sealed; it requires a drastic procedure of outgassing before filling; depending on the neutron wavelength, a high pressure of $^3$He, typically between 3 and 15 bar, is needed to provide adequate detection efficiency; regulation imposes to certify the gas vessel according to European standards. $CF_4$ or $Ar-CO_2$ is added to $^3$He to minimize the range of the ions, and to allow gas amplification. The pressure of this additive gas is determined according to the requirement on the spatial resolution independently from the efficiency requirement. For example, 3 bar of $CF_4$ is necessary to reduce the sole contribution of the gas to 1 mm FWHM. 2.5 times more pressure is needed with $Ar-CO_2$ (90-10) to reach the same resolution. For moderate spatial resolution and high counting rate applications, $Ar-CO_2$ is the favored gas compared to $CF_4$ because of a lower ageing effect. The necessity to use a high pressure of stopping/quenching gas (both functions are generally linked) to reach high position resolution comes from the fact that the Center Of Gravity of the ion track does not correspond to the neutron interaction point, and most of the electronics used for neutron detectors are based on the determination of the COG. This situation, specific to neutrons, imposes harsh limits to the detector: in particular, $^3$He detectors with resolution bellow 1 mm are difficult to build as they need a very high pressure of stopping/quenching gas, an electrode pitch smaller than the sought resolution, a high voltage on the electrodes, and a high signal/noise ratio. A resolution of 0.6 mm FWHM has been measured in 2D with a MSGC detector filled with 7 bar of $CF_4$ added to the $^3$He convertor gas [10]. Another important parameter is the precision by which the center of gravity of a Bragg peak or a reflection beam can be measured; this is typically one order of magnitude lower than the resolution FWHM needed. For beam monitors, the center of gravity of the beam must be measured with an accuracy of less than 1 mm, requiring a lower stopping gas pressure; a deviation of 0.7 mm RMS from the real value of a thermal neutron beam position has been obtained with a Borated multi layer GEM detector at ORNL test beam, as shown during the workshop; the detector was operated at the atmospheric pressure [11].

3D reconstruction of the ions tracks using TPC methods have been applied for high resolution 3He detectors operated at atmospheric pressure. It is necessary to recognize each of the two emitted particles, proton and tritium, in order to determine the neutron interaction point. This method requires complex and time-consuming algorithms which have not proved yet to be reliable enough for neutron detectors. The determination of the neutron interaction point by this method is very much simplified if the convertor is made of a thin film instead of a gas; in this case, the position of the neutron is at the intersection between the ion track and the convertor plan. By using a GEM or Micromegas detector, with the convertor film deposited on the anode, one can use the first active signal to determine the position of the neutron capture, as it has been done for example in neutronography with a Micromegas detector, or, if the count rate requirement is acceptable, it is possible to reconstruct the track in TPC mode to determine more accurately the intersection with the convertor film. If the convertor film is deposited on a cathode upstream the amplifying electrode, the position to be measured corresponds to the end of the track signal. The position measurement is more delicate in this case, but resolution half the readout pitch has already been demonstrated in this condition.

Another specificity of thermal neutron gas detectors is the high energy deposited in the gas after the reaction capture (0.76 MeV for $^3$He, and 2.31(94%) & 2.79 (6%) MeV for $^{10}$B), and the high density of electron-ion pairs along the track. It has some important consequences:
- The amplification gain needed to operate the detector is typically between 10 and 1000; this moderate value reduces the probability of destructive sparks, or make the detector more tolerant to fabrication imperfections.
- Compared to ions, Compton electrons produce tracks with a lower density charge. The rejection of gamma rays by threshold discrimination is an efficient method.

For a given amplification gain, the higher is the pressure of the stopping/quenching gas and the higher is the voltage on the anodes. The maximum gain a proportional counter can handle is limited by the maximum voltage it can sustain without sparking. It follows that the maximum signal/noise ratio decreases with the pressure. Both parameters, signal/noise ratio and stopping power, play in opposite direction for the spatial resolution. Designing high resolution $^3$He detectors is a matter of finding the best compromise between the signal/noise ratio and the stopping power of the gas.

As we said, replacing $^3$He by thin convertor films allows reaching high position resolution with the detector filled at atmospheric pressure. Another advantage of using thin convertor films is the fact that the detection gas can be flushed in the vessel, so the vessel doesn't need to be ultra-high vacuum compatible. This makes the fabrication of this type of detector much simpler than $^3$He detectors. Of course, all gas detectors, including MSGCs, MWPCs, and proportional counters benefit from this advantage.

One problem of thin convertor films is that the thickness needed to stop thermal neutrons (typically 30 micrometers for $^{10}$B) is much larger than the range of the particles produced by nuclear capture. Neutron captured at a depth superior than the alpha range will not produce charges in the gas, and can't be detected. Different techniques have been tested to tackle this problem. Examples of a few of them are listed below:

Coating the convertor films on micro-structured cathodes allows increasing the virtual thickness of B or $B_4C$ per unit of surface, hence the neutron absorption probability. A prototype with several $B_4C$ coated micro-structured cathodes combined with MWPCs has been tested at FRM2 [18,19]. Some detectors have been tested by the INFN Frascati-Bicocca collaboration using triple GEMs with $^{10}$B single blade up to 32 blades [11], as shown during the workshop. In those tests the detectors reached up to 31% efficiency at 5 meV with a gamma rejection of $10^{-7}$.

The CASCADE [12] detector is made of several (up to 10) GEMs, coated on both sides with $B_4C$ films of 1 micrometer and stacked together. A gain of 1 is applied to all GEMs, except to the last one which is operated at a higher gain. The last stage of this detector is made of the readout electrode. The CASCADE detector has demonstrated high count rate capability, and moderate spatial resolution. A complete characterization of its performances, in particular the neutron detection efficiency, the gamma rejection rate, the spatial uniformity, and the stability over time, as well as the background rate induced by scattered neutrons in the GEMs, is necessary to further explore the potential of this technique.

In a Multi-blade detector [13], the convertor films are oriented at grazing angle with respect to the incident neutrons. This is the macroscopic version of the micro-structured cathode. Several prototypes have been realized by different groups, with MWPC mounted between the blades; promising results have been obtained, in particular detection efficiency>30%, and high resolution in one direction. Nevertheless a lot of work remains to be done to demonstrate performances for neutrons scattering instruments, in particular, the uniformity of the detector, and its sensitivity with the neutron direction might create harsh difficulties to calibrate it. Given the configuration of this detector, it would be very attractive to use self-sustaining elements consisting of the convertor film, the amplifying electrode, and the 2D readout in one single plate. One might consider the use of thick-GEM for this application.

Ageing effect of detectors is a growing concern in neutron scattering science. This is due to the continuous increase of flux coming from progress with the source luminosity, and with the reflectivity of the neutron guides. Ageing effect is well quantified in high energy physics for continuous gas flushing; it does not depend on the type of particle to be detected, MIP or neutrons, but on the total charge current flowing through the anode wire. Some additional constraints exist for 3He detectors using sealed vessels: the detector must be outgassed between 80°C and 150°C during several weeks to remove molecules in particular of water adsorbed in the internal elements. Gas tightness is provided by metallic gaskets instead of organic materials; only low-outgassing materials, like ceramics, glass and metals are allowed; polyimide (Kapton©) connectors can replace ceramics connectors for reasons of simplicity, but there are queries about possible ageing induced by this material in sealed detectors.

## The effects of the $^3$He shortage crisis

$^3$He PSDs (Position Sensitive Detector) were commercially very successful until the $^3$He shortage crisis started in 2008. During many years a few companies (GE-Reuter Stokes, Centronic, Toshiba) fabricated $^3$He proportional counters for homeland security and oil well logging applications. These counters have also been used for NSS for long time until their progressive replacement by linear PSD

produced by a few companies, in particular Reuter Stokes and Toshiba. The market of $^3$He PSDs for NSS was beneficial thanks to the convergence of several factors:

- PSDs allows to satisfy the requirements of almost all instruments in NSS having a spatial resolution >= 8 mm FWHM. This is the strongest market in NSS in terms of surface of detectors.
- The equipment to produce proportional counters for the industry and PSDs for material science are the same.
- The technique of production is relatively simple, although it requires specific procedures based on company's know-how.
- These detectors are strongly reliable.

With the $^3$He shortage crisis threatening several fields of science and industry, it has been soon recognized as of prime importance by the Department Of Energy in US to develop alternative techniques whenever it was possible. Nowadays, $^3$He is available for neutron scattering Science but only in small quantities. Its price has reached 3000 €/l but it recently drop to 1500 €/l.

The high cost of $^3$He is acceptable for small and medium size detectors but not for large area detectors. In 2010, the neutron scattering scientific community identified three alternative neutron convertors to be developed in priority: $^{10}BF_3$ gas and $^{10}B$ thin films in gas detectors, and Scintillators coupled to photon counting devices.

## Detector requirements in neutron scattering science today

In order to perform high quality science, neutron facilities are involved in a continuous development of their instrument where detectors represent a key component. In Europe, ILL in France, LENA and TRIGA in Italy and FRM2 in Germany are using a reactor source, and ISIS in the UK and nTOF at CERN are using a spallation source. There is no fundamental difference between these two types of neutron sources in terms of instruments and detectors specifications. For neutron scattering instruments, the design of the instrument is all about transporting the "right part" of phase space from the neutron source to the sample under test. As such the instrument design can be highly varied, and so the detector requirements vary enormously between class of instrument design. Furthermore, the conditions on an instrument varying from one experiment to another, the detector must be optimized for the most frequent experimental condition, and be acceptable for extreme conditions. Large neutron facilities generally host suites of similar instruments specialized in one type of experiments, differing slightly from one to the other to cover the maximum range of experiments. As a result, standardization of detectors is often not possible, making transfer to the industry, and serial production difficult. The detector is one of many components of an instrument, each one with a technical risk. One challenge is to optimize the performance of the detector for each instrument with an acceptable technical risk. In most cases, there is no solution ready on the shelf in the industry; a compromise must be found between several parameters like cost, performance, technical risk, and reliability. Some examples are given to illustrate how these parameters interact together, but the subject exceeds by far the purpose of this summary.

**Efficiency $\varepsilon_c$ of the neutron convertor**

This parameter is the probability for a neutron going through the converting material to be captured and to produce a detectable signal. For a given neutron wavelength $\lambda$, it varies with the macroscopic cross section C of the convertor according to $\varepsilon_c = 1-e^{-C}$.

For thermal neutrons, C is proportional to $\lambda$. For short $\lambda$ values, a high $^3$He pressure is used to compensate the lower C value; this is simply done with 3He proportional counters by increasing the $^3$He pressure without changing anything else. For gas detectors using convertor films, the number (and cost) of films is increased. Gas detectors based on thin films, are competitive with 3He detectors only if the film convertor elements are significantly cheaper than 3He for equivalent macroscopic cross section.

**Efficiency $\varepsilon_e$ of the electronics and Gamma sensitivity**

In order to minimize the background coming from electronic noise and gamma events, a discrimination threshold $V_{th}$ is applied to the charge produced in the gas for validation. A gamma sensitivity less or equal to 10**-7 is usually recommended, but for several applications, 10**-5 is acceptable. The adequate value depends on the gamma background on the instrument, and on the type of instrument. For Time-Of-Flight instruments, a large fraction of the gamma produced downstream the beam chopper are suppressed by time rejection, hence $V_{th}$ can be reduced in this case to maximize $\varepsilon_e$

The quality of the electronics readout scheme, and the proper choice of the gas can help to minimize $V_{th}$ in order to reach maximum efficiency. This is particularly important for detectors which do not exhibit a clear separation between gammas and neutrons signal heights as it is the case for thin film convertors and scintillators.

**Sensitive Area (SA) and number of acquisition (Nacq)**

For neutron diffraction, several acquisitions might be necessary with different positions of the detector, or different orientations of the sample, to acquire a complete set of data. If the detector is made of several elements, the dead zones between them should be minimized.

**Time resolution and Sensitive Area**

For TOF instruments, the detector must be at a distance from the sample large enough to reach a good precision on the velocity measurement. Since a large angular coverage is also requested, this type of instrument is the most demanding in terms of sensitive area (several tens of m$^2$).

The time resolution required to measure the TOF is of the order of 10 microseconds, a value accessible to most of the counting detectors. Although there is no immediate benefit for neutron scattering science to use detectors with sub-microsecond resolution, time resolution of the order of 5 ns for the MPGDs can be exploited for fast neutrons experiments like n-TOF at CERN.

**Spatial resolution**

To compensate the lack of beam intensity, samples used in neutron science have generally a bigger volume than for Synchrotron radiation science. The spatial resolution required for the detector is usually expressed as the dimension of the smallest sample to be used on the instrument; it is typically 0.1 mm for protein crystallography with Laue diffractometers, 1 mm for reflectometers, or powder and single crystal diffractometers, and 1 cm for Inelastic scattering spectrometers. Except for Laue diffractometers, the state of the art of neutron gas detectors allows reaching these values for most of the current instruments, but higher resolution detectors are needed for ESS.

**Counting rate capability**

For applications requiring a large counting dynamics, in particular Reflectometry and Small Angle Neutron Scattering, it is often necessary to attenuate the neutron beam to avoid detector saturation. This situation, which is hardly acceptable regarding the investment to increase the luminosity of the neutron beams, will become critical on several ESS instruments.

**Counting stability and detector reliability**

These parameters might strongly influence the quality of an instrument. In almost all cases a detector with a high reliability and stability will be preferred to a promising detector with potentially better performances. Some applications require a relative counting variation less than $10^{-4}$. The detector should of course be reliable enough to be operational without interruption during several years. The ESS environment will push forward the requirement on radiation hardness.

## Applications in neutron scattering science

Following is a short list of applications in neutron scattering science where the availability of detectors is, or has been an important issue, and which received support from several European projects.

1) <u>Inelastic Instruments</u> (IN) are based on the measurement of the energy transfer of neutrons interacting in a sample. This is done by measuring the Time-Of-Flight (TOF) of neutrons between the chopper and the detector. The detector must cover a large solid angle, requiring a sensitive area of several tens of $m^2$. A position resolution of 1-2 cm is needed in both directions. The gamma sensitivity should not exceed $10^{-7}$ and the detection efficiency must be as high as possible, of the order of 70% for thermal neutrons. Position sensitive $^3$He tubes of 2-3 m length mounted on a cylinder with the sample position at the center is the currently preferred solution. The detectors are generally used in a large vacuum vessel to avoid scattering of neutrons in air. $^3$He is not available in quantity sufficient for this category of instrument, hence developing solutions free of $^3$He for inelastic instruments has been one of the highest priorities of neutron large scale facilities. One task in the CRISP European project (2011 – 2014) is to develop a detector demonstrator based on the Multi-Grid detector[14], which includes 15 or more layers of proportional counter tubes coated with $B_4C$ [17]. Results obtained recently show that this technique offers a robust and efficient solution to cover large sensitive area, with the cost of the detector dominated by the cost of the $B_4C$ coating. Micro-structured substrates have been studied to increase the efficiency of the $B_4C$ layer in the FP7-2 NMI3 project (2012 - 2016) [18,19] and in the INFN R&D studies.

2) <u>Small Angle Neutron Scattering Instruments</u> need detectors of medium size (between 50 cm x 50 cm and 100 cm x 100 cm), with a spatial resolution of ~5 mm in both directions, and a global counting rate of $10^6$-$10^7$ Hz. Position Sensitive Tubes for SANS have been studied during the TECHNI FP5 European project (2000 – 2004); they have progressively replaced MWPC very successfully on most of the SANS instruments in large neutron facilities; they are cheaper, faster, and easier to maintain. The volume of $^3$He needed is typically 50 liters for a detector of 1 $m^2$. It is an acceptable solution for ESS, provided the limited availability of $^3$He will remain at the same level. Otherwise the MPGDs will be one of the alternative solutions, but dedicated studies should be addressed from now on for the high spatial resolutions in large area detectors, that require high density Front End Electronics with good radiation hardness performances.

3) <u>Reflectometers</u> need compact detectors (typically 20 cm x 20 cm) with a spatial resolution of 1 mm in one direction, 5 mm in the other direction, and a high local counting rate ($10^5$-$10^6$ $Hz/mm^2$). The small volume of $^3$He needed (10-20 liter.bar) can be considered as manageable; MWPCs developed during the MILAND JRA in the NMI3 FP6 project, or Position Sensitive Tubes, provide sufficient spatial resolution but they are missing a factor 10 or 100 in counting rate for ESS. Several MPGD detectors have been developed with success and are considered as promising solutions for reflectometers, in particular the GSPC-MSGC developed in the "detector JRA" within the NMI3 FP7-1 project (2009 – 2013). This detector is based on the detection with a matrix of PMTs of the light produced by a MSGC during the avalanche process in the gas, providing 0.6 mm resolution FWHM and 50 ns dead time.

The need for more performing detectors in Reflectometry has been given again high priority in the future SINE2020 project which will start in 2015: several techniques will be studied, based on different type of amplifying electrodes and different types of neutron convertors: 6Li scintillators readout with PMTs or SiPMT, either directly, or via optical fibers, $^3$He MSGCs with parallel charge division readout, RPCs and Micromegas with $B_4C$. Also GEM based detectors represent a valid option for this kind of application.

4) <u>Single crystal and powder diffractometers</u> use $^3$He curved detectors with a radius of curvature of typically 120°. MWPCs and MSGCs are being used with equally high success. These detectors are complex; they require qualified people for the fabrication and the maintenance, conditions which can only be found in the largest neutron facilities.

5) <u>Beam monitors</u> are of 2 types: the most demanded are quasi-transparent detectors mounted permanently on the beam lines to correct data for flux variations on the sample. Position information is not requested, but they must be very stable (relative variation < $10^{-4}$) and have a linear counting response. The detection efficiency must be sufficient to minimize statistical fluctuations but only a negligible fraction of the beam should be attenuated. In order to minimize perturbation in guiding neutrons to the sample, transparent beam monitors should be as compact as possible (max 2 cm in the beam direction). One difficulty is to put the exact quantity of convertor material for the sought efficiency; for very low efficiency beam monitors ($10^{-6}$ – $10^{-8}$), Nitrogen is a better choice than $^{3}$He thanks to its lower capture cross-section. Ar used as stopping gas should be avoided because of its activation under the high neutron flux ($10^{10}$ n/cm$^{2}$.s)

Solid film based on $^{nat}$Gd, $^{6}$Li, $^{10}$B, and $^{235}$U have also been used for beam monitors in thermal neutron scattering. In particular, $^{235}$U fission chambers provides very high signals, hence gas amplification is not necessary; this type of detector is insensitive to gammas, and very resistant against ageing. Another type of beam monitors is to control the beam profile at the end of a guide section. Scintillators coupled to CCD cameras are often used as imaging beam monitors. Quantification of the flux is usually not requested, but it is accessible by adding a transparent monitor upstream the imaging monitor. GEM-based fast and thermal neutron beam monitors have also been used to characterize the ISIS neutron beam, showing a counting rate stability of about 95% [8, 9, 15]

## Instrumental requirements for ESS

The ESS (European Spallation Source) will be the most intense source of neutrons for material sciences. For the 22 reference instruments for the ESS, as defined in the ESS technical design report, the detector requirements are shown in the table below. These requirements are beyond state-of-the-art in terms of present day sources, however are representative of the challenge for neutron detector development over the next decade in the field of thermal neutron scattering. It should be noted that all detectors require a 2D position measurement of the neutron; and all resolutions quoted are for each axis. In addition, due to the timing characteristics of the pulsed spallation source, and the typical associated phase space gymnastics, a certain resolution on the timing is required. This timing resolution is a convolution of the timing resolution of the detection process with the flight path uncertainty between the sample and the detector. Typically the time of the neutron detection ("event") must be related to the energy by the time-of-flight method, as in the nuclear conversion information on the neutron energy is lost.

| Instrument | Detector Area [m²] | Wavelength Range [Å] | Time Resolution [µs] | Resolution [mm] |
|---|---|---|---|---|
| Multi-Purpose Imaging | 0.5 | 1-20 | 1 | 0.001 - 0.5 |
| General Purpose Polarised SANS | 5 | 4-20 | 100 | 10 |
| Broad-Band Small Sample SANS | 14 | 2-20 | 100 | 1 |
| Surface Scattering | 5 | 4-20 | 100 | 10 |
| Horizontal Reflectometer | 0.5 | 5-30 | 100 | 1 |
| Vertical Reflectometer | 0.5 | 5-30 | 100 | 1 |
| Thermal Powder Diffractometer | 20 | 0.6-6 | <10 | 2x2 |
| Bi-Spectral Powder Diffractometer | 20 | 0.8-10 | <10 | 2.5x2.5 |
| Pulsed Monochromatic Powder Diffractometer | 4 | 0.6-5 | <100 | 2 x 5 |
| Material Science & Engineering Diffractometer | 10 | 0.5-5 | 10 | 2 |
| Extreme Conditions Instrument | 10 | 1-10 | <10 | 3x5 |
| Single Crystal Magnetism Diffractometer | 6 | 0.8-10 | 100 | 2.5x2.5 |
| Macromolecular Diffractometer | 1 | 1.5-3.3 | 1000 | 0.2 |
| Cold Chopper Spectrometer | 80 | 1-20 | 10 | 10 |
| Bi-Spectral Chopper Spectrometer | 50 | 0.8-20 | 10 | 10 |
| Thermal Chopper Spectrometer | 50 | 0.6-4 | 10 | 10 |
| Cold Crystal-Analyser Spectrometer | 1 | 2-8 | <10 | 5-10 |
| Vibrational Spectroscopy | 1 | 0.4-5 | <10 | 10 |
| Backscattering Spectrometer | 0.3 | 2-8 | <10 | 10 |
| High-Resolution Spin Echo | 0.3 | 4-25 | 100 | 10 |
| Wide-Angle Spin Echo | 3 | 2-15 | 100 | 10 |
| Fundamental & Particle Physics | 0.5 | 5-30 | 1 | 0.1 |
| **Total** | **282.6** | | | |

Table 2.5: Estimated detector requirements for the 22 reference instruments in terms of detector area, typical wavelength range of measurements and desired spatial and time resolution.

In addition to these requirements, detection efficiency is important. Due to the predominance of a single technology in neutron scattering until very recently, there is a commonly held belief that this is more-or-less unity for most Helium detectors. Whilst this may not be completely the case, it is important that future detectors have a high efficiency; detection efficiencies such as 30% in the thermal region, to over 50% in the cold region can be seen as minimal efficiencies that are acceptable. Signal to background is of course the relevant parameter from the data; therefore backgrounds must be minimal. How much this is a deciding factor depends upon the exact application; for imaging a couple of orders of magnitude in dynamic range may be sufficient, but for inelastic processes, many of the signals being sought are hidden in the noise, and dynamic ranges of $10^6$ or beyond are desirable. In all cases however, a good gamma and fast neutron rejection should be sought; scattering of neutrons must be minimized by reducing dead material; and the inherent noise of the detector should be negligible.

Lastly, as the primary users of the neutron instruments are not experts, the detectors must "just work" and be transparent to the users, seamlessly integrated into the data-taking. This is a non-trivial requirement, and imposes restrictions on the lifetime and stability of the detectors. Any calibration of the detectors should be invisible to users of the facility. In this sense, progress has been made recently; a thermal neutron GEM based detectors has been successfully interfaced with the standard ISIS Data Acquisition Electronics (DAE), known as DAE2. This gave the possibility to perform a neutron diffraction measurement using the same DAQ as standard $^3$He detectors [16].

Technologies foreseen for realizing the baseline ESS instruments as outlined in the Technical Design Report are shown in the table below. In terms of identifying a niche where micro-pattern detectors may play a role, there are two aspects which play to their strengths; firstly in terms of spatial resolution, secondly in terms of rate requirements. In terms of spatial resolution, the instrument classes that wish for higher spatial resolution than presently achievable may benefit from novel developments in MPGD technologies for neutron detection. Secondly, because the local instantaneous rate at ESS will be an order of magnitude than at present sources, there is a wide range of instruments that may benefit from novel developments.

| Instrument | $^{10}$B Thin Films ⊥ | $^{10}$B Thin Films ∥ | Scintillators WLS | Scintillators Anger | $^3$He | Micropattern Rate | Micropattern Resolution |
|---|---|---|---|---|---|---|---|
| Multi-Purpose Imaging | - | - | - | - | - | o | + |
| General Purpose Polarised SANS | o | + | - | + | o | + | - |
| Broad-Band Small-Sample SANS | o | + | - | + | - | + | - |
| Surface Scattering | o | + | - | + | o | + | - |
| Horizontal Reflectometer | - | o | - | + | + | o | - |
| Vertical Reflectometer | - | o | - | + | + | o | - |
| Thermal Powder Diffractometer | o | + | + | - | - | o | - |
| Bi-Spectral Powder Diffractometer | o | + | + | - | - | o | - |
| P-M Powder Diffractometer | o | + | + | - | - | o | - |
| MS Engineering Diffractometer | o | + | + | - | - | o | - |
| Extreme Conditions Diffractometer | o | + | + | - | - | o | - |
| Single Crystal Diffractometer | o | + | + | - | - | o | - |
| Macromolecular Diffractometer | - | o | o | o | - | + | + |
| Cold Chopper Spectrometer | + | o | o | - | - | - | - |
| Bi-Spectral Chopper Spectrometer | + | + | o | - | - | - | - |
| Thermal Chopper Spectrometer | + | + | + | - | - | - | - |
| Cold Crystal Analyser Spectrometer | - | o | - | + | + | - | - |
| Vibrational Spectrometer | - | o | - | o | + | - | - |
| Backscattering Spectrometer | - | o | - | + | + | - | - |
| High-Resolution Spin Echo | - | o | - | o | + | + | - |
| Wide-Angle Spin Echo | - | o | - | o | + | + | - |
| Fundamental & Particle Physics | - | - | - | - | + | + | + |

## Defining Measurement Standards for Thermal Neutron Detectors

Measurement standards are a hot topic for neutron detection. As the neutron is destroyed in the detection event, it is difficult to unambiguously define that the detected signal was definitely from a neutron – eg a triggered setup that might be used to determine efficiencies to other particle types (such as in a cosmic or test beam telescope) is not possible here, as the detectors need to be self-triggered.

Neutron detection also is highly dependent upon the wavelength (energy) of the neutron; therefore a good quality beam of known and characterized wavelength is highly preferable. A typical neutron source, moderated with appropriate quantities of hydrogenous material (such as water or polythene, etc), will give a continuum of energies, and is therefore not a good source of neutrons for a detailed characterization of the detectors.

What is typically measured in terms of neutron detection is the relative efficiency to a "known, almost-black" Helium-3 device. This number is then turned into an absolute number by considering the efficiency of this device. This will probably give a good comparison between detectors measured on the same setup, however systematic effects are likely to be observed in terms of measuring devices between different setups. To eliminate these effects, which may be significant, a multi-layer setup, such as the 7-tube device used at the ILL, should be considered seriously, as by fitting the data from this device, an absolute efficiency number can be derived.

The boundary effects are important in considering detector efficiency, and should not be neglected. How the detector active area should be considered is application dependent. Similarly, the dead material in the detector device is important. Note that neutron transparency through various materials is very different to that of gamma's or particles as used in high-energy physics.

Whilst considering detector efficiency, it is important to ensure that this efficiency doesn't inadvertently contain gamma- or noise-induced events. It is also important that the same operational point is used to compare detector efficiency to the gamma efficiency.

Lastly, coherent and in-coherent scattering (ie diffraction and scattering) from the detector should be considered – this may erode significantly the dynamic range performance of the detector. Measuring this is application dependent and difficult to define in an abstract way.

In all the above, the ensure understanding of detector performance, the role of both analytical calculations and more detailed simulations (such as using GEANT or McSTAS) should not be underestimated. They will indicate whether the experimental results observed are reasonable.

Presently there are no widely accepted standards or universally-accepted procedures for measuring the above. In the context of the design of the European Spallation Source, presently existing and widely used methods will be collected and used to start to define these quantities in a self-consistent manner.

## Conclusions

Neutron detectors for scattering science must cope with a varying range of instrumental conditions:
- The sensitive area can be a few $cm^2$ up to several tens of $m^2$
- Most of the detectors need 2D position sensitivity
- The geometry required is typically planar, however for some applications, a cylindrical or spherical detector design would be advantageous
- A resolution of 0.1 mm FWHM is needed for protein crystallography and neutron radiography, whereas 2 cm FWHM is sufficient for inelastic scattering instruments
- The ability to determine the center of gravity of a neutron spot should be a small fraction of the position resolution, typically 10% of the resolution FWHM.

- The global counting rate ranges from <1 Hz for Ultra-Cold neutrons detectors dedicated to fundamental physics up to >1 MHz for Small Angle Neutron Scattering and reflectometers.
- Low counting rate applications are sensitive to the background noise produced by fast neutrons, and gamma rays.
- Counting Stability over time is often crucial, requiring relative counting variation as low as $10^{-4}$.

Standard gas detectors, like MWPC and Proportional counters, and scintillator detectors are broadly used but they are limited by the rate capability. Up to now MPGDs have been adopted only on a limited number of instruments. In particular, $^3$He-filled MSGCs have demonstrated a high reliability and the capability to be operated with 7 bar of CF4 to achieve sub-millimeter resolution FWHM, but the development and the maintenance of these detectors require equipment and a technical expertise which exist only in one or two neutron facilities. The know how on MPGDs construction and maintenance is improving rapidly as demonstrated by several projects of detector upgrade, both in HEP and industrial and medical applications world wide.

Replacement of $^3$He by alternative convertors is a priority for large area detectors. The Multi-Grid detector based on $10B_4C$ thin films is developed since 2010 by ILL and ESS. Experimental results have lead us to the conclusion that Multi-Grid detectors can be produced in a reproducible way with performances acceptable compared to $^3$He detectors; we consider that this technique is the most attractive solution today for ESS to replace $^3$He in large area detectors.

In terms of development, one urgent need for ESS, and for all neutron facilities, is small size detectors (0.5 m$^2$) for reflectometers, having a resolution of 1 mm in 2 directions, a counting rate of $10^5$/mm$^2$.s, a $10^{-5}$ gamma sensitivity, and an efficiency of >50% for thermal neutrons (25 meV). A very important quality criteria is the stability/reliability of the detector. This development has been defined as one of the priorities in the future SINE2020 European project devoted to neutron instrumentation, and starting in 2015. We believe that such a detector could serve as a driving horse for the development of high count rate, high resolution neutron detectors for scattering science. Scintillator and gas detectors, in particular MSGCs, GEMs and Micromegas could play an important rule to push forward the technical limits of neutron scattering instruments. One of the intrinsic advantages of MPGDs is their ability to sustain high counting rate and to provide high position resolution. Standardization of methods to evaluate performances of neutron detectors is needed to take into account the interdependence of the detection parameters